\documentclass[prd,aps,twocolumn,preprintnumbers,nofootinbib,showpacs,amsmath,amssymb,superscriptaddress]{revtex4-1}

\usepackage{graphicx}
\usepackage{dcolumn}
\usepackage{bm}
\usepackage{hyperref}
\usepackage{epsfig}
\usepackage{cancel}
\usepackage{enumitem}
\usepackage[utf8]{inputenc}
\hypersetup{
    colorlinks=true,
    linkcolor=blue,
    filecolor=red,
    urlcolor=blue,
    citecolor=red
} 

\begin{document}

\title{Holographic entanglement negativity conjecture for adjacent intervals in 
$\mathrm{AdS}_3/\mathrm{CFT}_2$ }

\author{Parul Jain}
\email{parul.jain@ca.infn.it}
\affiliation{Dipartimento di Fisica, Universit\`a di Cagliari\\Cittadella Universitaria, 09042 Monserrato, Italy}
\affiliation{ INFN, Sezione di Cagliari, Italy}

\author{Vinay Malvimat}
\email{ vinaymm@iitk.ac.in }
\affiliation{Department of Physics,\\
Indian Institute of Technology Kanpur,\\
Kanpur 208016, INDIA}
\author{Sayid Mondal}
\email{ sayidphy@iitk.ac.in}
\affiliation{Department of Physics,\\
Indian Institute of Technology Kanpur,\\
Kanpur 208016, INDIA}
\author{Gautam Sengupta}
\email{sengupta@iitk.ac.in}
\affiliation{Department of Physics,\\
Indian Institute of Technology Kanpur,\\
Kanpur 208016, INDIA}

\begin{abstract}
\noindent

We propose a holographic entanglement negativity conjecture involving the bulk geometry, for mixed states of adjacent intervals in $(1+1)$-dimensional dual conformal field theories through the $AdS/CFT$ correspondence. The holographic entanglement negativity is obtained from a specific algebraic sum of the geodesics anchored on respective intervals on the boundary which reduces to the holographic mutual information between them. Utilizing our conjecture
we obtain the entanglement negativity of adjacent intervals in zero and finite temperature ($1+1$)-dimensional holographic conformal field theories dual to the bulk $AdS_3$ vacuum and the Euclidean BTZ black hole respectively. Our holographic conjecture exactly reproduces the conformal field theory results obtained through the replica technique, in the large central charge limit.  We briefly elucidate the corresponding issue for the $AdS_{d+1}/CFT_d$ scenario.

\end{abstract}

\maketitle
\section{Introduction}
\label{sec1}

Quantum entanglement in recent times has impacted an expansive list of theoretical issues from condensed matter physics to quantum gravity through the holographic $AdS/CFT$ correspondence \cite{VanRaamsdonk:2009ar,VanRaamsdonk:2010pw,PhysRevD.86.065007,Maldacena:2013xja,Hartman:2013qma}. This geometric connection has allowed the characterization of quantum entanglement in extended systems like holographic conformal field theories. For bipartite quantum systems in a pure state this involves the {\it entanglement entropy} which is defined as the von Neumann entropy of the reduced density matrix. In a series of interesting communications Calabrese {\it et al} advanced a comprehensive procedure to compute the entanglement entropy of $(1+1)$-dimensional conformal field theories ($CFT_{1+1}$) \cite{Calabrese:2004eu,Calabrese:2009qy} utilizing the replica technique. 

Following \cite{Calabrese:2004eu}, in a seminal work Ryu and Takayanagi proposed a holographic characterization of the entanglement entropy in $d$-dimensional conformal field theories $(CFT_d)$, involving bulk dual $AdS_{d+1}$ geometries through the $AdS/CFT$ correspondence \cite{Ryu:2006bv,Ryu:2006ef} ( for an extensive review see \cite {Nishioka:2009un}). According to the Ryu and Takayanagi ( RT) conjecture the universal part of the entanglement entropy of a subsystem in a dual $CFT_d$ was described by the area of a co-dimension two bulk $AdS_{d+1}$ static minimal surface homologous to the subsystem. For the
$AdS_3/CFT_2$ scenario the static minimal surface reduces to a space like geodesic in the bulk $AdS_3$ geometry anchored on the appropriate spatial interval in the dual $CFT_{1+1}$.
The holographic entanglement entropy obtained from the RT conjecture exactly reproduces the corresponding $CFT_{1+1}$ results obtained through the replica technique in the large central 
charge limit.


It is well known however in quantum information theory that the entanglement entropy ceases to be a valid measure for the characterization of mixed state entanglement where it receives contributions from irrelevant correlations. This is a complex issue in quantum information theory and necessitates the introduction of suitable entanglement measures for the {\it distillable} entanglement. In a seminal communication Vidal and Werner \cite{PhysRevA.65.032314} addressed this issue and proposed a computable measure termed {\it entanglement negativity} which provides an upper bound on the {\it distillable entanglement} whose non convexity was subsequently demonstrated by Plenio in \cite {Plenio:2005cwa}. In the recent past Calabrese {\it et al} in \cite {Calabrese:2012ew,Calabrese:2012nk,Calabrese:2014yza} utilized an alternative replica technique to compute the entanglement negativity for mixed states in both zero and finite temperature $CFT_{1+1}$. 

Naturally the above developments lead to the significant issue of a holographic description for the entanglement negativity involving the bulk geometry for dual conformal field theories in the $AdS/CFT$ frame work \cite {Rangamani:2014ywa}. In a recent interesting communication two of the present authors (VM and GS) in the collaborations \cite {Chaturvedi:2016rcn,Chaturvedi:2016rft,Chaturvedi:2016opa}(CMS),  proposed a holographic conjecture for the entanglement negativity of mixed states in holographic $CFT_d$s. The conjecture involves a specific  algebraic sum of the areas of co-dimension two bulk extremal surfaces ( geodesic lengths in $AdS_3/CFT_2$ ) anchored on the corresponding subsystems. Note that the proposed conjecture was in relation to the entanglement negativity for a singly connected subsystem in an infinite system. Remarkably, this conjecture precisely characterizes the upper bound on  the {\it distillable entanglement} through the elimination of the thermal contributions for finite temperature mixed states. Furthermore it could be shown that the holographic entanglement negativity was proportional  to the sum of the holographic mutual information for the relevant partitioning of the system.

In this article motivated by the {\it CMS conjecture}, \cite{Chaturvedi:2016rcn,Chaturvedi:2016rft,Chaturvedi:2016opa} we advance an independent holographic entanglement negativity conjecture for the mixed states of adjacent intervals in both 
zero and finite temperature $CFT_{1+1}$. In the context of the $AdS_3/CFT_2$ scenario we establish that the holographic entanglement negativity for this configuration is described by a specific algebraic sum of the bulk geodesic lengths which is proportional to the holographic mutual information between the intervals. Note that our case is distinct from \cite {Hartman:2013mia} where the entanglement entropy characterizing the entanglement between the intervals and the rest of the system was considered. On the other hand here we consider the entanglement negativity which characterizes the entanglement between these adjacent intervals. Our computation for the entanglement negativity is expected to have significant applications to various entanglement issues in diverse fields.

This article is organized as follows. In section \ref{sec2} we briefly review the 
characterization of entanglement negativity in quantum  information theory and describe the computation of this quantity for mixed states of adjacent intervals in a $CFT_{1+1}$. In the subsequent section \ref{sec4} we establish our holographic conjecture for the entanglement negativity which exactly reproduces the replica technique results for the $CFT_{1+1}$ in the large central charge limit. We then summarize our results in the last section \ref{sec5}.


\section{Entanglement Negativity}
\label{sec2}

In this section we review the characterization of entanglement negativity in quantum information theory and the computation of this entanglement measure for mixed state configuration of adjacent intervals in a $CFT_{1+1}$ for both zero and finite temperatures
as well as for a finite size system.

\subsection{Entanglement negativity in quantum information theory}
We begin by briefly  reviewing the definition of entanglement negativity in quantum information theory as described in \cite{PhysRevA.65.032314,Plenio:2005cwa}. In this context it is required to consider a tri-partition of a quantum system into the subsystems $A_1$, $A_2$ and $B$, with $A=A_1\cup A_2$ and $B=A^c$ representing the rest of the system. The Hilbert space $\mathcal{H}$ for the bipartite system $A$ may be written as a direct product  $\mathcal{H}=\mathcal{H}_1 \otimes \mathcal{H}_2$ of the corresponding Hilbert spaces of the individual subsystems $A_1$ and $A_2$ respectively. The reduced density matrix of the subsystem $A$ is given as $\rho_A=\mathrm{Tr}_{A^c}(\rho)$. The partial transpose of this reduced density matrix with respect to $A_2$ may be defined as follows
\begin{equation}\label{26}
\langle e^{(1)}_ie^{(2)}_j|\rho_A^{T_2}|e^{(1)}_ke^{(2)}_l\rangle = 
\langle e^{(1)}_ie^{(2)}_l|\rho_A|e^{(1)}_ke^{(2)}_j\rangle,
\end{equation}
where $|e^{(1)}_i\rangle$ and $|e^{(2)}_j\rangle$ are the bases for the respective Hilbert spaces
$\mathcal{H}_1$ and  $\mathcal{H}_2$.
The entanglement negativity between the subsystems $A_1$ and $A_2$ is then described as
follows
\begin{equation}\label{28}
\mathcal{E} = \ln \mathrm{Tr}|\rho_A^{T_2}|,
\end{equation}
where the trace norm $\mathrm{Tr}|\rho_A^{T_2}|$ is given by the sum of absolute eigenvalues of the partially transposed reduced density matrix $\rho_A^{T_2}$. 

\subsection{Entanglement negativity in $CFT_{1+1}$}
\begin{figure}
\begin{center}
\includegraphics[scale=1.5,keepaspectratio]{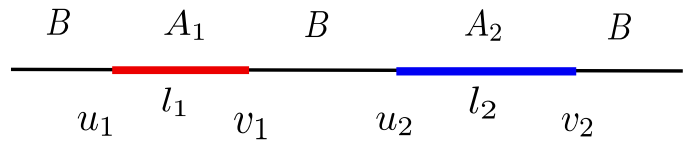}
\caption{Schematic of two disjoint intervals $A_1$ and $A_2$ in a $(1+1)$-dimensional ${CFT}$.}\label{fig1}
\end{center}
\end{figure}

Here we briefly review the computation of the entanglement negativity for mixed states of two disjoint intervals in a $CFT_{1+1}$ as described in \cite{Calabrese:2012ew,Calabrese:2012nk}. To this end the authors considered the tri-partition of the system as depicted in Fig. (\ref{fig1}), where
$A_1$ describes the interval ($A_1\in[u_1,v_1]$) of length $l_1$ and $A_2$ is the interval ($A_2\in[u_2,v_2]$) of length $l_2$ while $B=A^c$ represents the rest of the system.
The entanglement negativity between the subsystems $A_1$ and $A_2$ is obtained through a suitable replica technique which involves the construction of the quantity $\mathrm{Tr}(\rho_A^{T_2})^{n_e}$ for even $n_e$ and considering the analytic continuation of the even sequence to $n_e\to1$ as \footnote{Note that the explicit construction of this analytic continuation is complex and remains an open issue except for some simple conformal field theories \cite{Calabrese:2009ez, Calabrese:2010he,Calabrese:2014yza} (See also \cite{Headrick:2010zt}).} 
\begin{equation}\label{29}
\mathcal{E} = \lim_{n_e \rightarrow 1 } \ln \mathrm{Tr}(\rho_A^{T_2})^{n_e}.
\end{equation}
It was shown that the quantity $\mathrm{Tr}(\rho_A^{T_2})^{n_e}$ in the above expression may be expressed as a four point function of the twist fields on the complex plane as follows
\begin{equation}\label{61}
\mathrm{Tr}(\rho_A^{T_2})^{n_e} = 
\langle\mathcal{T}_{n_e}(u_1)\overline{\mathcal{T}}_{n_e}(v_1)\overline{\mathcal{T}}_{n_e}(u_2)\mathcal{T}_{n_e}(v_2)\rangle_{\mathbb{C}}.
\end{equation}

\subsection{Adjacent intervals in vacuum}\label{adj_int_vac}
\begin{figure}
\begin{center}
\includegraphics[scale=1.5,keepaspectratio]{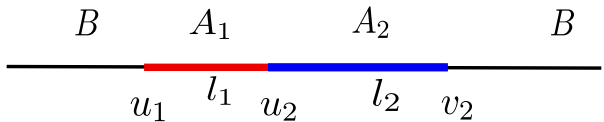}
\caption{Schematic of two adjacent intervals $A_1$ and $A_2$ in a $(1+1)$-dimensional ${CFT}$.}\label{fig2}
\end{center}
\end{figure}

The formulation described above was utilized to characterize the 
entanglement negativity for mixed states of adjacent intervals in \cite{Calabrese:2012ew,Calabrese:2012nk}. This configuration is obtained through the limit $v_1\rightarrow u_2$ for Fig. (\ref{fig1}) which reduces to the configuration depicted in Fig. (\ref{fig2}). 

The four point function eq. (\ref {61}) in this limit effectively reduces to a three point function of the twist fields as follows
\begin{equation}\label{32}
\mathrm{Tr}(\rho_A^{T_2})^{n_e} = 
\langle\mathcal{T}_{n_e}(-l_1)\overline{\mathcal{T}}^2_{n_e}(0)\mathcal{T}_{n_e}(l_2)\rangle, 
\end{equation}
The entanglement negativity for the adjacent intervals at zero temperature is obtained by employing eqs. (\ref{29}) and (\ref{32}), which may be expressed as
\begin{equation}\label{35}
\mathcal{E} = \frac{c}{4} \ln \bigg[\frac{l_1l_2}{(l_1+l_2)a}\bigg]+ const,  
\end{equation}
where $a$ and $c$ are the UV cut off and the central charge of the corresponding $CFT$. Note that the first term in the above expression is universal whereas the
second `const' term depends on the detailed operator content of the $CFT$.

\subsection{Adjacent intervals in vacuum for finite size}\label{adj_int_vac_fin_size}
\label{sec2.4}

For a finite sized system of length $L$ with a periodic boundary condition, the entanglement negativity for the adjacent intervals may be computed from the corresponding 
three point function  eq. (\ref{32}) on a cylinder\cite{Calabrese:2012nk}. This is obtained through the conformal map 
$z \rightarrow w = \frac{iL}{2\pi}\ln z$, from the complex plane to the cylinder with the circumference 
$L$. The entanglement negativity may then be expressed as
\begin{equation}\label{19}
\mathcal{E} =
\frac{c}{4} \ln \bigg[\Big(\frac{L}{\pi a}\Big)\frac{\sin(\frac{\pi l_1}{L})\sin(\frac{\pi l_2}{L})}
{\sin\frac{\pi (l_1+l_2)}{L}}\bigg] +const.
\end{equation}
As earlier the `const' term in the above expression is non universal.

\subsection{Adjacent intervals at finite temperature}
An analysis as above may be performed to obtain the entanglement negativity for the mixed state of adjacent intervals in a finite temperature $CFT_{1+1}$ which is defined on
a cylinder of circumference $\beta=\frac {1}{T}$ which is the inverse temperature. This is related to the $CFT_{1+1}$ on the complex plane through the conformal 
map $z \rightarrow w = \frac{\beta}{2\pi}\ln z$.
Using the above transformation and  eq. \eqref{29}, the entanglement negativity is given as
\begin{equation}\label{22}
\mathcal{E} =
\frac{c}{4} \ln \bigg[\Big(\frac{\beta}{\pi a}\Big)\frac{\sinh(\frac{\pi l_1}{\beta})\sinh(\frac{\pi l_2}{\beta})}
{\sinh\frac{\pi (l_1+l_2)}{\beta}}\bigg] +const, 
\end{equation}
where the `const' as earlier is non universal.
\subsection{Large central charge limit}\label{lar_c}
Here we review the large central charge analysis for the entanglement negativity of the mixed state configuration of adjacent intervals at zero temperature as described in \cite{Kulaxizi:2014nma}. According to the $AdS_3/CFT_2$ correspondence, the holographic limit in which the bulk spacetime is semiclassical ( Newton's constant $G_N^{(3)}\to0$) is equivalent to the large central charge limit $c\to \infty$ in the dual $CFT_{1+1}$ as these quantities are related through the Brown-Henneaux formula $c=\frac{3 R}{2 G_N^{(3)}}$. In \cite{Kulaxizi:2014nma}, such a large central charge analysis for the four point correlator in eq. (\ref{61}) was obtained by employing the monodromy technique to determine the conformal blocks providing the dominant contribution. The authors demonstrated that in the limit $v_1\rightarrow u_2$, which corresponds to the case of adjacent intervals depicted in Fig. (\ref{fig2}),  the dominant contribution to the four point function in eq. (\ref{61}) arises from the conformal block corresponding to the intermediate operator $\overline{{\mathcal{T}}}^2$. This led  to the following expression for the large central charge limit of the entanglement negativity as
\begin{equation}\label{enlargec}
{\cal E}=\frac{c}{4} \ln \Big[\frac{l_1l_2}{(l_1+l_2)a}\Big].
\end{equation}
Note that this matches exactly with the universal part of the entanglement negativity of the adjacent intervals given in eq. (\ref{35}), implying that the non-universal constant term in eq. (\ref{threepoint}) is sub leading in the large central charge limit\footnote { Note that the monodromy technique was also utilized earlier to determine the large central charge limit of the entanglement entropy for multiple disjoint intervals in \cite{Hartman:2013mia} (and through a different technique in \cite{Headrick:2010zt}). The authors demonstrated that in this limit the entanglement entropy is universal and matched exactly with the predictions of the holographic Ryu-Takayanagi conjecture.}. The authors also demonstrated numerically that the entanglement negativity vanishes ${\cal E}\to 0$ in the limit of disjoint intervals placed far apart. In the subsequent sections we will establish a holographic conjecture which exactly reproduces the large-c results for the case of the adjacent intervals from the corresponding dual bulk geometry.
\section{Holographic entanglement negativity for adjacent intervals}
\label{sec4}

In this section we establish our holographic entanglement negativity conjecture involving the bulk $AdS_3$ geometry for the mixed state configuration of adjacent intervals in a dual $CFT_{1+1}$. As described in (\ref {lar_c}) the large central charge analysis for the entanglement negativity clearly indicates the plausibility of a holographic characterization for the entanglement negativity in a dual $CFT_{1+1}$ through the $AdS_3/CFT_2$ correspondence. To this end we consider the adjacent intervals $A_1$ and $A_2$ of lengths $l_1$ and $l_2$ respectively as depicted in Fig. (\ref{fig2}), where the subsystem $A_1\cup A_2$ is in a mixed state. As described in (\ref {adj_int_vac}) for this mixed state configuration in the $CFT_{1+1}$, the entanglement negativity involves the three point twist correlator in eq. \eqref {32} whose form is fixed from conformal invariance as follows
\begin{equation}\label{threepoint}
\begin{split}
&\langle\mathcal{T}_{n_e}(z_1)\overline{\mathcal{T}}_{n_e}^{2}(z_2)\mathcal{T}_{n_e}(z_3)\rangle_{\mathbb{C}} \\&~~~~~=  
\frac{c_{n_e}^2~C_{\mathcal{T}_{n_e}\overline{\mathcal{T}}_{n_e}^2\mathcal{T}_{n_e}}}{|z_{12}|^{\Delta_{\mathcal{T}^2_{n_e}}}
|z_{23}|^{\Delta_{\mathcal{T}^2_{n_e}}}|z_{13}|^{2\Delta_{\mathcal{T}_{n_e}}-\Delta_{{\mathcal{T}}^2_{n_e}}}},
\end{split}
\end{equation}
where $c_n$ are normalization constants and $C_{\mathcal{T}_{n_e}\overline{\mathcal{T}}_{n_e}^2\mathcal{T}_{n_e}}$ is the structure constant for the relevant twist field OPEs. The scaling dimensions $\Delta_{\mathcal{T}_{n_e}}$ and 
$\Delta_{\mathcal{T}_{n_e}^2}$ of the twist fields $\mathcal{T}_{n_e}$ and $\mathcal{T}_{n_e}^2$ are respectively given as \cite {Calabrese:2012nk}
\begin{equation}\label{34}
\begin{aligned}
\Delta_{\mathcal{T}_{n_e}} =  \frac{c}{12}\Big(n_e - \frac{1}{n_e}\Big),\\
\Delta_{\mathcal{T}_{n_e}^2} =2\Delta_{\mathcal{T}_{\frac{n_e}{2}}}= \frac{c}{6}\Big(\frac{n_e}{2}- \frac{2}{n_e}\Big).
\end{aligned}
\end{equation}

In this connection we note here that the two point twist correlators in the $CFT_{1+1}$ are given as
as follows \cite {Calabrese:2012ew,Calabrese:2012nk}
\begin{equation}\label{twoptfn}
\begin{split}
\big<{\cal T}_{n_e}(z_k)\overline{{\cal T}}_{n_e}(z_l)\big>_{\mathbb{C}}=\frac{C_{}^{(1)}}{|z_{kl}|^{\Delta_{\mathcal{T}_{n_e}}}},\\
\big<{\cal T}^2_{n_e}(z_k)\overline{{\cal T}}^2_{n_e}(z_l)\big>_{\mathbb{C}}=\frac{C_{}^{(2)}}{|z_{kl}|^{\Delta_{\mathcal{T}^2_{n_e}}}},
\end{split}
\end{equation}
where $C^{(1)},~ C^{(2)}$ are normalization constants.
A close examination of the form of the above two point twist correlator now clearly indicates that the three point twist correlator in eq. \eqref{threepoint} may be factorized as follows 
\begin{equation}\label{factor}
\begin{split}
&\langle\mathcal{T}_{n_e}(z_1)\overline{\mathcal{T}}_{n_e}^{2}(z_2)\mathcal{T}_{n_e}(z_3)\rangle_{\mathbb{C}}=C~\big<{\cal T}_{n_e}(z_1)\overline{{\cal T}}_{n_e}(z_3)\big>_{\mathbb{C}}\\
&\bigg(\frac{\big<{\cal T}^2_{{n_e}{}}(z_1)\overline{{\cal T}}^2_{{n_e}{}}(z_2)\big>_{\mathbb{C}}\big<{\cal T}^2_{{n_e}{}}(z_2)\overline{{\cal T}}^2_{{n_e}{}}(z_3)\big>_{\mathbb{C}}}{\big<{\cal T}^2_{{n_e}{}}(z_1)\overline{{\cal T}}^2_{{n_e}{}}(z_3)\big>_{\mathbb{C}}}\bigg)^{1/2},
\end{split}
\end{equation}
where the constant $C$ is given as follows
\begin{equation}
C=c_{n_e}^2~C_{\mathcal{T}_{n_e}\overline{\mathcal{T}}_{n_e}^2\mathcal{T}_{n_e}}C_{}^{(1)}({ C_{}^{(2)}})^{1/2}.
\end{equation}
Interestingly, the leading universal part of the three point function eq. \eqref{factor}, which is dominant in the large central charge limit may be expressed as follows
\begin{equation}\label{3ptfactin2pt}
\begin{split}
&\langle\mathcal{T}_{n_e}(z_1)\overline{\mathcal{T}}_{n_e}^{2}(z_2)\mathcal{T}_{n_e}(z_3)\rangle_{\mathbb{C}}=C~\big<{\cal T}_{n_e}(z_1)\overline{{\cal T}}_{n_e}(z_3)\big>_{\mathbb{C}}\\&~~~~~~\frac{\big<{\cal T}_{\frac{n_e}{2}}(z_1)\overline{{\cal T}}_{\frac{n_e}{2}}(z_2)\big>_{\mathbb{C}}\big<{\cal T}_{\frac{n_e}{2}}(z_2)\overline{{\cal T}}_{\frac{n_e}{2}}(z_3)\big>_{\mathbb{C}}}{\big<{\cal T}_{\frac{n_e}{2}}(z_1)\overline{{\cal T}}_{\frac{n_e}{2}}(z_3)\big>_{\mathbb{C}}}.
\end{split}
\end{equation}
Note that in the above expression in eq. (\ref{3ptfactin2pt}) we have used the following factorization in a $CFT_{1+1}$ as described in \cite{Calabrese:2012ew,Calabrese:2012nk}
\begin{equation}
 \langle{\cal T}^2_{n_e}(z_i)\overline{{\cal T}}^2_{n_e}(z_j)\rangle_{\mathbb{C}}=\big<{\cal T}_{\frac{n_e}{2}}(z_i)\overline{{\cal T}}_{\frac{n_e}{2}}(z_j)\big>^2_{\mathbb{C}} .
\end{equation}
We have ignored the non-universal constants $c_n$ and $C_{\mathcal{T}_{n_e}\overline{\mathcal{T}}_{n_e}^2\mathcal{T}_{n_e}}$  in eq. (\ref{factor}), in the large central charge limit as mentioned in (\ref {lar_c}).

It is well known from the $AdS_3/CFT_2$ correspondence that the two point function of the twist fields located at $z_i$ and $z_j$ in a $CFT_{1+1}$ is related to the length of the dual bulk space like geodesic $\mathcal{L}_{ij}$  anchored on the corresponding interval as follows \cite{Ryu:2006bv,Ryu:2006ef}
\begin{eqnarray}
&\big<{\cal T}_{n_e}(z_k)\overline{{\cal T}}_{n_e}(z_l)\big>_{\mathbb{C}} \sim e^{-\frac{\Delta_{n_e}{ \cal L}_{kl}}{R}},\label{4}\\
&\big<{\cal T}_{\frac{n_e}{2}}(z_i)\overline{{\cal T}}_{\frac{n_e}{2}}(z_j)\big>_{\mathbb{C}} \sim e^{-\frac{\Delta_{\frac{n_e}{2}}{ \cal L}_{ij}}{R}},~~~~~\label{5}
\end{eqnarray}
where $R$ is the $AdS_3$ radius. The three point twist correlator in eq. (\ref{factor}), upon employing eqs. \eqref{4} and \eqref{5} may be expressed as 
\begin{equation}\label{10}
\begin{split}
 &\langle\mathcal{T}_{n_e}(z_1)\overline{\mathcal{T}}_{n_e}^{2}(z_2)\mathcal{T}_{n_e}(z_3)\rangle_{\mathbb{C}}\\& ~~~
=\exp{\bigg[\frac{-\Delta_{\mathcal{T}_{n_e}}\mathcal{L}_{13}-\Delta_{\mathcal{T}_{\frac{n_e}{2}}}(\mathcal{L}_{12}+\mathcal{L}_{23}-\mathcal{L}_{13})}{R}\bigg]}.
\end{split}
\end{equation}
Here the corresponding points defining the adjacent intervals
are $(z_1=-l_1, z_2=0,z_3=l_2)$. In the replica limit\footnote{Note that the large central charge limit has to be taken prior to the replica limit to arrive at eq.(16). This order of limits is critical and is also true for the case of the entanglement entropy where the scaling dimension of the twist field 
${\cal \tau}_n$ vanishes in the replica limit and has to be understood in the sense of an analytic continuation (see \cite{Hartman:2013mia,Fitzpatrick:2014vua}). In a similar fashion the negative scaling dimension of the twist field ${\cal \tau}_{\frac{n_e}{2}}$ in the replica limit is in the sense of the non-trivial analytic continuation involved in the replica definition for the entanglement negativity.}($n_e\to 1$) the scaling dimensions $\Delta_{\mathcal{T}_{n_e}}\to 0$ and $\Delta_{\mathcal{T}_{\frac{n_e}{2}}}\to-\frac{c}{8}$. Hence by utilizing eqs. (\ref{10}) and (\ref{32}) in eq. (\ref{29}), the holographic entanglement negativity for the mixed state of the adjacent intervals in a dual $CFT_{1+1}$ may be expressed in terms of a specific algebraic sum of the lengths of the bulk space like geodesics $\mathcal{L}_{ij}~(i,j\in 1,2,3)$ anchored on appropriate intervals as follows
\begin{equation}\label{11}
\mathcal{E} = \frac{3}{16G^3_N}(\mathcal{L}_{12}+\mathcal{L}_{23}-\mathcal{L}_{13}), 
\end{equation}
where we have used the Brown-Henneaux formula ${c}=\frac{3R}{2G^3_N}$  \cite{brown1986}. Upon employing the Ryu-Takayanagi conjecture \cite{Ryu:2006bv} for the holographic entanglement entropy  the eq. (\ref{11}) above may be expressed   in terms of a specific algebraic sum of the holographic entanglement entropies as follows
 \begin{equation}\label{14}
\mathcal{E} =  \frac{3}{4}(S_{A_1}+S_{A_2}-S_{A_1\cup A_2})=\frac{3}{4}
{\cal I}(A_1, A_2).
\end{equation}
This algebraic sum of the holographic entanglement entropies is precisely the
{\it holographic mutual information} ${\cal I}(A_1,A_2)$  between the two intervals for the mixed state configuration under consideration.
\begin{figure}
\begin{center}
\includegraphics[scale=1.5,keepaspectratio]{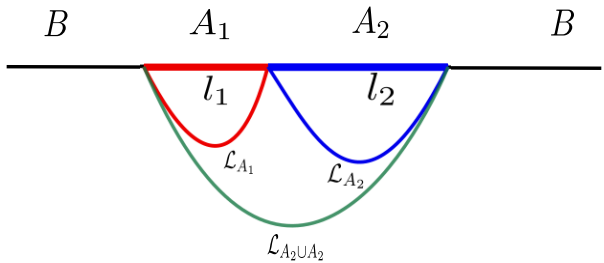}
\caption{Schematic of bulk geodesics anchored on the subsystems $A_1$, $A_2$ and $A_1\cup A_2$ in a $(1+1)$-dimensional boundary ${CFT}$.}\label{fig3}
\end{center}
\end{figure}
It is important to note that the mutual information and the entanglement negativity are entirely distinct entanglement measures in quantum information theory. Negativity is an upper bound on the distillable entanglement and the mutual information is the upper bound on the total correlation in the bipartite system. However the universal parts of these two quantities which are dominant in the holographic (large-$c$) limit are proportional for the particular mixed state configuration,
although this is not the case for other configurations where a more intricate relation between these quantities are valid in the holographic limit \footnote{ Interestingly this proportionality has also been observed in the context of global and local quench in a $CFT_{1+1}$ \cite{Coser:2014gsa,Wen:2015qwa} for the same configuration utilizing the replica technique. However note that our results are distinct and pertain to a holographic description ( distinct from the replica technique) of entanglement negativity
for the mixed state of a dual $CFT_{1+1}$ without quench.}.

As explained in the introduction the entanglement negativity in our case characterizes the entanglement between the adjacent intervals for the mixed state configuration of a $CFT_{1+1}$ in question. It is important to mention here that eq. \eqref{11} provide an elegant holographic construction involving the dual bulk $AdS_3$ geometry for the entanglement negativity of this mixed state in terms of a specific algebraic sum of the lengths of space like geodesics anchored on the intervals and their union. Interestingly as shown in eq. \eqref{14} this may be expressed as an algebraic sum of the corresponding entanglement entropies which reduces to the holographic mutual information for the mixed state configuration in question. Our construction characterizes a significant progress for the critical issue of a holographic description of entanglement negativity in a generic $AdS_{d+1}/CFT_d$ scenario started by two of the present authors in \cite {Chaturvedi:2016rcn}. The present work is an important sequel towards further elucidation of this crucial issue and provides critical insight into the structure of holographic entanglement negativity for generic mixed states in $CFT_d$s. Our construction for the mixed state configuration under consideration constitutes a significant example 
for the holographic description of entanglement negativity and indicates 
an elegant geometrical technique which may generalized for application to other mixed state configurations in a $CFT_{1+1}$.

The construction described here suggests a possible higher dimensional generalization of the holographic negativity conjecture for mixed states of adjacent subsystems in a generic $AdS_{d+1}/CFT_d$ scenario. The entanglement negativity in this case would involve a similar algebraic sum of the areas of bulk co-dimension two static minimal surfaces anchored on the appropriate subsystems, which may be expressed as follows
\begin{equation}\label{HEN CONJ AREA}
\mathcal{E} = \frac{3}{16G^{(d+1)}_N}\big(\mathcal{A}_{1}+\mathcal{A}_{2}-\mathcal{A}_{12}\big)=\frac{3}{4}
{\cal I}(A_1, A_2),  
\end{equation}
where $\mathcal{A}_{i}$ denotes the area of the bulk co dimension two static minimal surface
anchored on the subsystem $A_i$. Note that the above expression is also proportional to the holographic mutual information between the two subsystems. However the higher dimensional generalization requires substantiation through specific examples and a possible proof from a bulk perspective which constitute an open issue \footnote { Note that the higher dimensional holographic entanglement negativity construction suggested above has been implemented for mixed state configurations of adjacent subsystems described by rectangular strip geometries in $CFT_d$s dual to bulk pure $AdS_{d+1}$ space time, $AdS_{d+1}$-Schwarzschild and RN-$AdS_{d+1}$ black holes in \cite {Jain:2017xsu,Jain:2018bai}. The results reproduce certain universal features of entanglement negativity for $CFT_{1+1}$s which are consistent with quantum information theory expectations and constitute significant consistency checks for the higher dimensional conjecture.}.

\subsection{Adjacent intervals in the vacuum}

The holographic conjecture established above in eq. (\ref {11}) which describes the entanglement negativity for the mixed state configuration of adjacent intervals in terms of the lengths of the bulk space like geodesics anchored on appropriate intervals may now be applied to specific examples. In this context we first obtain the holographic entanglement negativity for the mixed state configuration of adjacent intervals at zero temperature in a $CFT_{1+1}$ dual to a bulk pure $AdS_3$ geometry. The bulk metric in the Poincar$\mathrm{\acute{e}}$ coordinates is given as
\begin{equation}\label{50}
 ds^2 = -\left(\frac{r^2}{R^2}\right)dt^2 +\left(\frac{r^2}{R^2}\right)^ {-1} dr^2 + \frac{r^2}{R^2} dx^2,
\end{equation}
where $R$ is the $AdS_{3}$ radius and the coordinate $x \in \mathbb{R} $. The bulk spacelike geodesic length ${\cal L}_{\gamma}$ anchored to the interval $\gamma$ of length $l_{\gamma}$  in the $CFT_{1+1}$ is given as follows \cite{Ryu:2006bv, Ryu:2006ef,Cadoni:2009tk,e12112244}
\begin{equation}\label{54}
\mathcal{L}_{\gamma} = 2R \ln \frac{l_{\gamma}}{a}, 
\end{equation}
where $a$ is the UV cut off. The holographic entanglement negativity for the adjacent intervals of lengths $l_1$ and $l_2$ may now be expressed
by employing  eq. \eqref{11} as follows
\begin{equation}\label{53}
\mathcal{E} = \frac{c}{4} \ln \bigg[\frac{l_1l_2}{(l_1+l_2)a}\bigg],
\end{equation}
where we have used the Brown-Henneaux formula.
Remarkably, the holographic entanglement negativity eq. (\ref{53}) matches  exactly with the corresponding replica technique result in eq. \eqref {35} at the large central charge limit.

\subsection{Adjacent intervals for finite sized systems in vacuum}

We now proceed to compute the holographic entanglement negativity for mixed states of adjacent intervals in a finite sized system of length $L$ at zero temperature in a dual $CFT_{1+1}$ from our conjecture described in eq. \ref{11}. In this case as explained in (\ref{adj_int_vac_fin_size}) the corresponding $CFT_{1+1}$ is defined on an infinite cylinder 
with the spatial direction compactified on a circle of circumference $L$. The bulk dual configuration for this case is the $AdS_3$ vacuum in global coordinates with the following metric \cite{Ryu:2006bv, Ryu:2006ef,Cadoni:2009tk,e12112244} 
\begin{equation}\label{51}
ds^2= R^2(-\cosh^2 \rho dt^2 + d\rho^2 + \sinh^2  \rho d \phi^2 ),
\end{equation}
where $\phi$ is periodic with a period of $2\pi$. The length of a bulk space like geodesic anchored on the subsystem 
$\gamma$ of length $l_\gamma$ in the 
boundary $CFT_{1+1}$ is given as \cite{Ryu:2006bv, Ryu:2006ef,Cadoni:2009tk,e12112244}
\begin{equation}\label{55}
\mathcal{L}_{\gamma} = 2R \ln \Big(\frac{L}{\pi a}\sin \frac{\pi l_{\gamma}}{L}\Big). 
\end{equation}
The holographic entanglement negativity for the adjacent intervals of lengths $l_1$ and $l_2$ may now be computed by substituting the respective geodesic lengths given by eq. \eqref{55} in eq. \eqref{11}, which leads to the following expression
\begin{equation}\label{56}
\mathcal{E} = \frac{c}{4} \ln \bigg[\Big(\frac{L}{\pi a}\Big)\frac{\sin(\frac{\pi l_1}{L})\sin(\frac{\pi l_2}{L})}
{\sin\frac{\pi (l_1+l_2)}{L}}\bigg],
\end{equation}
where we have utilized the the Brown-Henneaux formula. Interestingly, the holographic entanglement negativity obtained in eq. (\ref{56}) exactly reproduces the replica technique results eq. (\ref{35})  in the large central charge limit. 
\subsection{Adjacent intervals at finite temperature}
We now proceed to compute the holographic entanglement negativity for the mixed state of 
adjacent intervals in a dual $CFT_{1+1}$ at a finite temperature utilizing our conjecture described in eq. (\ref{11}). The corresponding $CFT_{1+1}$ is defined on a spatially infinite cylinder with the Euclidean time direction compactified on a circle of circumference $\beta$.
The appropriate bulk dual configuration is then described by a Euclidean BTZ black hole \cite{Ryu:2006bv, Ryu:2006ef,Cadoni:2009tk,e12112244} 
at a Hawking temperature $T=\frac{1}{\beta}$ with the $\phi$ direction uncompactified\footnote {Note that this configuration is actually a BTZ black string.}. The metric for this is given as
\begin{equation}\label{52} 
ds^2 = \frac{\left(r^2 - r_{h}^2\right)}{R^2}d\tau^2 +\frac{R^2}{\left(r^2 - r_{h}^2\right)}dr^2 + \frac{r^2}{R^2}d\phi^2 , 
\end{equation}
where $r=r_h$ denotes the event horizon and $\tau$ is the Euclidean time.
The length of a bulk spacelike geodesic anchored on a subsystem $\gamma$ is given as \cite{Ryu:2006bv,Ryu:2006ef,Cadoni:2009tk,e12112244}
\begin{equation}\label{57}
\mathcal{L}_{\gamma} = 2R \ln \Big(\frac{\beta}{\pi a}\sinh \frac{\pi l_{\gamma}}{\beta}\Big), 
\end{equation}
where $\beta=2\pi R^2/r_h$ is the inverse Hawking temperature. 

It is now possible to compute the holographic entanglement negativity for the mixed state of adjacent intervals of lengths $l_1$ and $l_2$ in the finite temperature $CFT_{1+1}$ from our conjecture by utilizing the eqs. \eqref{57} and \eqref{11}, which leads to the following expression
\begin{equation}\label{58}
\mathcal{E} = \frac{c}{4} \ln \bigg[\Big(\frac{\beta}{\pi a}\Big)\frac{\sinh(\frac{\pi l_1}{\beta})\sinh(\frac{\pi l_2}{\beta})}
{\sinh\frac{\pi (l_1+l_2)}{\beta}}\bigg],
\end{equation}
where once again we have employed the Brown-Henneaux formula. As earlier the holographic entanglement negativity eq. (\ref{58}) exactly reproduces the replica technique result eq. (\ref{22}) for the finite temperature  $CFT_{1+1}$ in the large central charge limit. This clearly constitutes compelling evidence for the validity of our holographic entanglement negativity conjecture.
\section{Summary and Discussion}
\label{sec5}

To summarize we have established a holographic   entanglement negativity conjecture for mixed states of adjacent intervals in zero and finite temperature dual $CFT_{1+1}$ 
in the $AdS_3/CFT_2$ scenario. Note that this configuration is a mixed state as the degrees of freedom of the rest of the system are traced over and hence the corresponding entanglement negativity characterizes the entanglement between the adjacent intervals. Our conjecture involves a specific algebraic sum of the lengths of bulk space like geodesics anchored on respective intervals in the dual holographic $CFT_{1+1}$ which reduces to the holographic mutual information between the intervals. 

In the $AdS_3/CFT_2$ scenario, the bulk dual configuration for the zero temperature $CFT_{1+1}$ is the pure $AdS_3$ vacuum expressed in terms of the Poincare coordinates for the infinite sized system and in the global coordinates for the finite sized system respectively. The finite temperature case on the other hand involves a bulk Euclidean BTZ black hole (black string). In both the cases we have clearly demonstrated that our conjecture exactly reproduces the corresponding replica technique results for both zero and finite temperature dual $CFT_{1+1}$s in the large central charge limit. Note however that our holographic conjecture in this article is specific to the mixed state configuration of adjacent intervals. From this perspective a more general holographic entanglement negativity conjecture for the mixed state of disjoint intervals is desirable. This constitutes an interesting open issue for a future investigation.

Naturally the holographic conjecture proposed here in the context of the $AdS_3/CFT_2$ scenario suggests a possible generalization in a $AdS_{d+1}/CFT_d$ scenario. The mixed state in this case would be defined by the configuration of adjacent subsystems in the dual holographic $CFT_d$. The holographic entanglement negativity would involve a specific algebraic sum of the areas of bulk co dimension two extremal surfaces anchored on the respective subsystems, which would be proportional to the holographic mutual information between them. However as such a higher dimensional generalization would require substantiation through specific examples and a possible proof from  a bulk perspective.

We emphasize here that our holographic conjecture provides a simple and elegant method to compute the entanglement negativity for such mixed states in holographic CFTs both at zero and finite temperatures. Naturally this is expected to have significant applications in diverse areas like strongly coupled theories in condensed matter physics, topological phases, quantum phase transitions and critical issues in quantum gravity. These are interesting open questions for the future.
\section{Acknowledgement}
Parul Jain would like to thank Prof. Mariano Cadoni for his guidance and 
the Department of Physics, Indian Institute of Technology Kanpur, India for their warm hospitality. Parul Jain's work is financially supported by Universit\`a di Cagliari, Italy and INFN, Sezione di Cagliari, Italy.
\bibliographystyle{utphys} 
\bibliography{HENparul}
\end{document}